\documentstyle[12pt]{article}

\makeatletter
\@addtoreset{equation}{section}
\makeatother

\begin{document}
\hfill{}
\hfill{}

\hfill{UCSBTH-97-14}

\hfill{hep-th/9706204}

\vspace{24pt}

\begin{center}
{\large {\bf Absorption of Scalars by Extended Objects}}

\vspace{48pt}

Roberto Emparan

\vspace{12pt}

{\sl Department of Physics}\\
{\sl University of California}\\
{\sl Santa Barbara, CA 93106, USA}\\
{\it emparan@cosmic1.physics.ucsb.edu}

\vspace{72pt}

{\bf Abstract}
\end{center}
\begin{quote}{\small
We derive the low energy absorption cross section of minimal scalars 
for a very generic class of geometries, that includes, among others, 
black holes, strings, $p$-branes, as well as intersecting brane 
configurations. The scalar field can be absorbed across a regular 
surface (a horizon) of finite area, or across a zero area throat, 
which may be singular or regular. Then we focus on some particular 
cases and compare with the results obtained using microscopic models 
at weak coupling. Exact agreement is found for the absorption by the 
non-degenerate ground state of effective strings. For geometries where
the throat is singular, we show that absorption through the stretched 
horizon yields correct results for degenerate states. For 
non-degenerate (ground) states we introduce a {\it stretched throat}, 
and show that it accounts for the correct dependence on frequency, 
charges, and moduli for the absorption by ground states of BPS 
fundamental strings. It is also shown to lead to the correct 
dependence on frequency for extremal D-branes. 
}\end{quote}

\newpage

\section{Introduction}

{}Following the successful identification of the microscopic stringy 
origin of the Bekenstein-Hawking entropy \cite{ello}, 
a great deal of effort 
has been put into comparing
details of the dynamics of weakly coupled strings to that of
black holes at larger coupling. 
In particular, the emission and 
absorption of scalars by near extremal black holes in five and
four dimensions have been shown to be remarkably well described
in terms of the tree-level interactions of modes of 
an effective string
\cite{callan,dhar,dmtwo,gubkleb1,ms,gubkleb2,fixeds,interm} 
\footnote{Partial disagreement has been found in some cases at some
regions of the near 
extremal parameter space \cite{km,fay,taylor,krakle}.}. 
The fact that the agreement of detailed radiation profiles
shown by Strominger and Maldacena in
\cite{ms} turns out to be exact is certainly remarkable. 
But what is 
most striking is the nature of the quantities 
that agree. For the black hole, greybody factors arise as a 
consequence of the gravitational potential barriers surrounding 
the horizon, i.e., an effect of spacetime curvature. In contrast,
their origin in the stringy description lies in the thermal 
distribution of excitations of the D-brane bound state. Together
with the Bekenstein-Hawking entropy, this is probably the strongest 
hint of a deep and mysterious connection between curvature and 
statistical mechanics. 

At present, the nature of such connection is far from clear.
But the possibility appears that by studying details of the classical
emission/absorption of scalars we can learn something about the 
dynamics of strings.
Conversely, by comparing the classical and 
string descriptions we hope to gain
some insight on the geometrical manifestations of
dynamics at microscopic scales.

Most of the work in these directions has been focused on 
the absorption/emission properties of extremal 
or near extremal black holes in $D=5$ or $D=4$ with regular horizons
of finite area. Nevertheless, outside
this domain a remarkable success 
has been achieved in \cite{threebr}, 
where it has been shown that the
absorption of scalars by extremal D-threebranes can be accurately
reproduced by worldvolume interactions at weak coupling. This is
particularly surprising if we consider that the degeneracy of
near-BPS excitations of threebranes, although it scales correctly
with the charges, does not reproduce precisely the 
numerical factor in the corresponding Bekenstein-Hawking 
entropy \cite{threent}. The absorption by 
the M-theory membrane and fivebrane has also been found 
in \cite{threebr} to yield
consistent results, 
although in these cases the approach is limited by the
lack of detailed knowledge of the worldvolume 
theories for several coincident M-branes. Further work 
in \cite{gkt} has sharpened the consistency
of the picture.

Progress along other lines has been slower. The microscopic
dynamics of dilatonic black holes, which become singular
in the extremal limit, is not sufficiently understood yet. 
In this context, Sen \cite{sen2} could 
correctly identify
the degeneracy of perturbative BPS string states with the area
of a stringy {\it stretched horizon} \footnote{This notion was 
introduced in a slightly different context in \cite{susskind}.}.
A broader framework capable of dealing with dilatonic black holes, 
and even black holes far from extremality, has been proposed 
in \cite{hp} \footnote{A different program is being
developed in \cite{cvelar}.}. 
A simple correspondence between perturbative strings
and black holes at a certain value of the coupling
seems to be enough to correctly match
the entropies (up to factors of order one). However, trouble has 
appeared when trying to compare absorption rates \cite{mio,das}, 
suggesting
that the dynamics may undergo significant changes along the transition
from the string to the black hole description \cite{mio,math}.

Apart from its implications for the connections between strings
and black holes, the study of semiclassical absorption rates 
has yielded results of interest on its own.
It has been possible to analyze with great generality the low
energy scalar absorption by black holes in arbitrary
dimensions \cite{dgm}. This has revealed a universal property: the 
absorption cross section of minimal scalars 
at vanishing frequencies always equals 
the area of the horizon. This means, in particular, that the 
absorption cross section for a zero area black hole should vanish 
at zero frequency. Nevertheless, and this is a main theme 
in this paper, in such cases one can
compute the leading term in the absorption cross section and extract
interesting information from it. This was indeed the case for the
absorption by non-dilatonic branes in \cite{threebr}.
The black holes that result from wrapping
such branes have singular horizons of zero area, even though the
geometry of the extended non-dilatonic
branes is non-singular. In fact, 
the absorption cross section of many
extended objects (such as extremal strings and $p$-branes
with $p\geq 1$), does
in general decrease to zero with the frequency due to
contraction of the geometry along the worldvolume directions.
But the surface that absorbs the scalar field need not 
have finite size to produce an absorption cross section of physical 
significance.

In this paper we extend the general analysis of absorption
by black holes in \cite{dgm} by considering scalars impinging
in directions transverse to an extended object. The latter 
can be almost any of the branes, as well as their 
intersections and boosts, that have been constructed in recent 
years as low energy solutions of string and M-theory.
One of our results is that when the geometry of the absorbing object 
is regular (even if it may have zero area), the absorption cross
section presents a simple form that extends the universality
properties of the scalar absorption at low energies.

The paper is structured as follows: in Sec.\ \ref{absor} we introduce
the class of geometries that we deal with, discuss the problem 
of scattering scalars at low energies off the absorption surfaces, 
and then compute
the s-wave absorption cross section. 
In Sec.\ \ref{anat} we
digress on the geometrical structure of the absorption surfaces 
(the throats) for several cases of particular interest. Then,
in Sec.\ \ref{wkly} we pick out some examples where a microscopic
description of the absorption can be formulated, and compare this
with the results of the semiclassical calculation. In particular, 
in \ref{bpsfs} we argue in favor
of the notion of a {\it stretched throat} (a regular geometry of
zero area) to account correctly for the absorption by states with
zero entropy (ground states). 
Sec.\ \ref{concl} summarizes the results. 
The calculation of the absorption of higher partial waves
is presented in an appendix.

\section{Absorption of low frequency s-wave scalars}\label{absor}

We will consider the propagation of a massless scalar field 
$\Phi$ minimally coupled to a geometry of the generic form
\footnote{We could use reparametrizations to arrange 
for metrics such that $f(r)=h(r)$. 
However, it is convenient for
applications to consider the form (\ref{metric}).}
\begin{equation}\label{metric}
ds^2 = \gamma_{\mu\nu}(r) dx^\mu dx^\nu + f(r) dr^2 + 
r^2 h(r) d\Omega_{n+1},
\end{equation}
Here $\gamma_{\mu\nu}$  $(\mu,\nu=0,\dots,p)$
is a metric on a $(p+1)$-dimensional
space ${\cal M}_p$ 
of Lorentzian signature, with
metric coefficients depending only on the transverse radial 
coordinate $r$. 
The geometry is assumed to be asymptotically flat,
\begin{eqnarray}
&&\gamma_{\mu\nu}(r)\rightarrow \eta_{\mu\nu},
\nonumber\\
&&f(r),\, h(r) \rightarrow 1,
\end{eqnarray}
as $r\rightarrow\infty$. 

Here we will only deal with s-waves transverse 
to ${\cal M}_p$, i.e., $\Phi= e^{-i\omega t}\phi_\omega (r)$. These 
dominate the absorption at low energies. The
study of higher partial waves 
is deferred to the appendix. Also, we could incorporate
momenta parallel to ${\cal M}_p$, but we will not do it here.

The wave equation now takes the form
\begin{equation}\label{waveq}
\left[ -\omega^2 \gamma^{tt}(r) f(r) +
{1\over \sqrt{-\gamma(r)\over f(r)}[r^2 h(r)]^{n+1\over 2}}
{d\over dr} \sqrt{-\gamma(r)\over f(r)}[r^2 h(r)]^{n+1\over 2}
{d\over dr}\right]\phi_\omega(r) =0.
\end{equation}
Here $\gamma = \det \gamma_{\mu\nu}$. The scattering problem is 
appropriately analyzed in terms of the ``tortoise'' coordinate $r^*$, 
defined by
\begin{equation}
dr^* =  dr\sqrt{-\gamma^{tt} f(r)}.
\end{equation}
{}Furthermore, it is convenient to introduce a function $U(r)$ as
\begin{equation}
U =\left[\gamma \gamma^{tt}[r^2 h(r)]^{n+1}\right]^{1/2},
\end{equation}
and substitute
\begin{equation}
\phi_\omega = {1\over \sqrt{U}} \psi.
\end{equation}
The wave equation can then be written as a Schr\"odinger equation
\begin{equation}\label{scheq}
\left[ -{d^2\over (dr^*)^2} + V\right]\psi = 
\omega^2 \psi
\end{equation}
with potential 
\begin{equation}\label{poten}
V = {1\over \sqrt{U}} {d^2 \sqrt{U}\over (dr^*)^2}.
\end{equation}

We are interested in the low energy ($\omega\rightarrow 0$)
absorption of the scalar across the surface spanned by ${\cal M}_p$
at $r=0$, which we will refer to as the 
``absorption
surface''. Typically, it will be a horizon 
(extreme or 
non-extreme) or a $p$-brane throat. Note, 
however, that in principle it need not have any such interpretation: 
$r=0$ is just
a surface where we impose boundary conditions for the field. 
Specifically, we will impose the boundary condition that, as we 
approach
$r=0$, the field contains only ingoing waves. The absorption 
probability is then obtained as the ratio of the flux ${\cal F}_0$
at $r=0$ to 
the incoming flux at infinity, ${\cal F}^{in}_\infty$, the
fluxes being given by
\begin{equation}\label{flux}
{\cal F} = {1\over 2i} 
{U\over \sqrt{-\gamma^{tt} f(r)}}\left(\phi_\omega^* 
{d\phi_\omega\over dr} - c.c.\right).
\end{equation}
The absorption cross section per unit 
volume of the transverse space ${\cal M}_p$ is then 
\begin{equation}
\sigma = {(2\pi)^{n+1}\over \omega^{n+1}\Omega_{n+1}} 
{{\cal F}_0\over {\cal F}^{in}_\infty},
\end{equation}
where the volume of the unit $(n+1)$-sphere is
\begin{equation}
\Omega_{n+1} = {2\pi^{n/2+1}\over \Gamma\left({n\over 2}+1\right)}.
\end{equation}
It is important to notice that the only information about the 
geometry that we need in order to compute 
the absorption cross section is contained in the two functions
$\sqrt{-\gamma^{tt} f(r)}$ and $U$.

At long distances,  $r^*\approx r$ and
$V\approx {n^2-1\over 4 r^2}$.
The wave equation is solved, as in a flat background,
in terms of Bessel functions,
\begin{equation}\label{farsol}
\phi_\omega^\infty (r) = {1\over (\omega r)^{n/2}} 
\left( A J_{n/2}(\omega r) +B J_{-n/2}(\omega r)\right),
\end{equation}
and the incoming flux is 
\begin{equation}\label{fluxinf}
{\cal F}^{in}_\infty = {|A +B|^2\over 2\pi \omega^n}.
\end{equation}

In order to solve the equation near the absorption surface we 
need to specify the 
behavior of the two relevant functions close
to $r=0$. We will be very generic and parametrize them 
in terms of 
two dimensionful parameters and two exponents as
\begin{eqnarray}\label{parexp}
U &\approx& S\; r^{a-b},\nonumber\\
 \sqrt{-\gamma^{tt}(r) f(r)} &\approx&  {T\over r^{b+1}}.
\end{eqnarray}
The dimensions of the parameters $S$ and $T$ are
$[{\rm length}]^{n+1-a+b}$ and $[{\rm length}]^{b+1}$, respectively, 
and they are typically related
to the mass and/or charges of a $p$-brane. 
The definition of the exponents
$a$ and $b$ has been adequately chosen for later convenience.

Now, notice that when $a=b$, the area of the absorption
surface is finite, and equal to 
$U|_{r=0}\;\Omega_{n+1}=S\;\Omega_{n+1}$ (times an unimportant
factor of the extent $L^p$ of the worldvolume coordinates). 
If $a>b$ this area
is zero, and it diverges if $a<b$. Further insight on the
qualitative features of the absorption surface can be gained
by studying the potential near $r=0$. This is  
\begin{equation}\label{nearpot}
V\approx (a^2- b^2){r^{2b}\over 4T^2}.
\end{equation}
As long as 
$|a|<|b|$ the potential is 
negative. As a consequence, all quanta with vanishingly low energies 
are absorbed. We are not interested in these cases. In 
contrast, for 
$|a|>|b|$ the scalar field has to penetrate through a potential 
barrier in order to reach to $r =0$, and therefore the absorption
at zero frequency will vanish. Finally, when $|a|=|b|$ the 
potential is vanishingly flat 
and the absorption cross section will tend
to a constant value at small frequencies.

When there are repulsive potential barriers, $|a|> |b|$, we can 
still distinguish three qualitatively different situations as to
their
dependence on $r$:
(a) if $b < 0$ the potential barriers become infinite as we approach 
$r=0$, and there is no absorption of quanta \footnote{
This is true in cases where the potential
is smooth enough. But since, for $b<0$, $r=0$ is at finite
``tortoise'' coordinate distance ($r^*=0$), one can possibly 
construct geometries with barriers that are thin enough to 
allow for penetration at finite frequencies. We will not
consider such situations.}; (b) for 
$b=0$ (and $a\neq 0$), the potential tends to a constant in the 
vicinity of $r =0$ ($r^*\rightarrow -\infty$) 
and the low energy quanta cannot, in general,
penetrate the barrier; (c) for $b>0$ absorption is possible since 
the potential
vanishes near $r=0$. In such situations 
the potential barrier for the massless s-wave scalar typically
starts at zero
near infinity, grows to a maximum as the radius decreases, and then 
approaches zero near $r=0$. For dilatonic black holes in four
dimensions, these cases (a), (b) and (c) 
correspond to dilaton 
coupling $a_d >1$, $a_d =1$, and $a_d < 1$, respectively,
and this sort of analysis was carried out
in \cite{holwil}.

{}Focus for the moment on situations where the absorption
surface is behind finite
potential barriers, i.e.,
\begin{equation}
a\geq b >0.
\end{equation} 
Near the absorption surface, the wave equation with potential
(\ref{nearpot}) is again a Bessel equation, and
the solution can be conveniently given
in terms of the first Hankel function $H^{(1)}_{\nu}$, 
\begin{equation}\label{nearsol}
\phi^{\rm near}_\omega  ={1\over (\omega r)^{a/2}}
H^{(1)}_{a\over 2b} \left({\omega T\over b r^b}\right),
\end{equation}
which contains only ingoing waves 
near $r=0$, and therefore satisfies the required boundary condition.
With the chosen normalization, the ingoing flux is
\begin{equation}\label{fluxo}
{\cal F}_0 = {2b\over \pi} {S\over T\omega^a}.
\end{equation}

The absorption cross section will be fully determined when we fix
the integration constants $A$ and $B$ in the far region solution 
(\ref{farsol}).
To do so, we have to match it to the near region solution
(\ref{nearsol}). This requires the existence of an overlap region 
where both solutions are approximately valid. Typically, for smooth
potentials and at low 
frequencies, the near solution is valid for $\omega r\ll 1$, whereas 
the far solution holds when $r$ is much bigger than some 
power of the gravitational length scale of the geometry (usually
some power of $S$ or $T$). The matching is then feasible if
$\omega$ is much smaller than this scale. 
However, in general, the possibility of matching 
has to be checked for each specific situation under study. 
We will assume in the
following that an overlap region exists. Then
it is always possible to effect the matching: the leading
term in $\phi^{\rm near}_\omega$ at large distances 
tends to a constant
(independent of $r$), and this behavior coincides with the limiting
form, for small $\omega r$, of $\phi^{\rm \infty}_\omega$ with 
$B=0$. Then $A$ can be determined as
\begin{equation}
A= -{i\over \pi} 2^{n/2}\Gamma\left({a\over 2b}\right)
\Gamma\left({n\over 2}+1\right)
\left( {2b\over \omega^{b+1}T}\right)^{a\over 2b}.
\end{equation}
The absorption cross section can finally be written as
\begin{equation}\label{theeq}
\sigma = {\pi\over \Gamma({a\over 2b})^2} \Omega_{n+1} 
S
\left({T\omega\over 2b}\right)^{{a\over 
b}-1}.
\end{equation}
This is the formula we sought.
As a particular case, it follows that when $a=b$ 
the absorption cross
section tends to a constant
\begin{equation}\label{reghor}
\sigma = \Omega_{n+1} S, \qquad (a=b),
\end{equation}
equal to the area of the absorption surface. In this way we have 
generalized the result in \cite{dgm} to absorption by higher 
dimensional objects. Indeed, whenever $a=b$ the equations can be 
solved following the steps of \cite{dgm}, and one can check 
that (\ref{reghor}) is valid even when $a=b\leq 0$. 

{}For the last remaining case where
the potential becomes zero near $r=0$, i.e., $a=-b$, our formula
above would yield a divergent result at vanishing
frequencies. However, the matching of solutions
apparently breaks down. We will not dwell on this situation, since
we have not found any interesting configuration 
with this sort of parameters. 

When $a >b>0$,
$\sigma$ has a non-trivial dependence on 
the frequency, and vanishes at $\omega =0$, as advertised above.
Our point here is that the leading term (\ref{theeq}) 
is nonetheless interesting.

\section{Anatomy of the throat}\label{anat}

The generic form of the metric in (\ref{metric}) covers a wide 
range of backgrounds of interest, including extreme and 
non-extreme supergravity $p$-branes and black holes, with one or 
several charges, or boosts along some world volume direction, 
as well intersections of $p$-branes, or combinations of all
these. 
Here we
will discuss the structure of the geometry near the 
absorption surface $r=0$ 
for some situations of particular interest.

We have already discussed the absorption surfaces with non-zero 
area, which are the ones that yield non vanishing cross sections 
at zero frequency. There are, though, interesting 
situations where the area of the absorption surface is zero.
Let us consider just the angular $(n+1)$-spheres as 
$r\rightarrow 0$. If 
$[r^2 h(r)]$ takes on a finite, non-zero value at $r=0$ then the 
angular sphere has a finite size and the geometry will be typically 
regular. In many cases, however, $[r^2 h(r)]_{r=0}=0$, and the 
absorption surface usually develops
a singularity. The cases where $[r^2 h(r)]_{r= 0}
\rightarrow \infty$
are most certainly pathological (e.g., for the supergravity 
$p$-branes below, this happens only for imaginary dilaton coupling)
and will not be considered in the following.

The discussion can be refined further by restricting to a more
specific class of geometries, such as those where the 
worldvolume ${\cal M}_p$ is homogeneous and the transverse space is 
isotropic.
We parametrize this, near $r=0$, as
\begin{equation}\label{isotr}
ds^2\approx \left({r\over\nu}\right)^{2\beta} (-dt^2 +dx^m dx^m)
+ \left({\mu\over r}\right)^{2\alpha} (dr^2 +r^2d\Omega_{n+1}),
\end{equation}
where $m=1,\dots, p$, $\alpha\leq 1$, and $\mu$, $\nu$ are two length
parameters that can be related to $S$ and $T$ of the previous
section. This geometry is characteristic of extremal $p$-branes
in $D=p+n+3$ spacetime dimensions.
The geometry near $r=0$ can be 
regarded as a `throat', down which the scalar field is absorbed.

Throats with $\alpha< 1$ are singular: the angular spheres
pinch down to zero size. Focus for the moment 
on $\alpha =1$.
In this case, the exponents $\alpha$, $\beta$ are related to those
defined in (\ref{parexp}) as $a=(p+1)\beta$ and $b=\beta$. 
These finite size throats
are all regular, and in fact the asymptotic metric is the product
of the angular sphere with
$(p+2)$-dimensional anti-de Sitter space
\cite{ght}. To see this, change the radial variable to 
\begin{equation}
y={|b|\over \mu}\left({r\over \nu}\right)^{b}
\end{equation}
to find
\begin{equation}
ds^2 \approx {\mu^2\over b^2}\left[ y^2 (-dt^2 +dx^m dx^m)
+ {dy^2\over y^2}\right] + \mu^2 d\Omega_{n+1}.
\end{equation}
This is $(adS)_{p+2}\times S^{n+1}$, a generalization of 
the Bertotti-Robinson geometry (which corresponds to $p=0$, $n=1$). 
If $b>0$, 
then $r\rightarrow 0$ corresponds to $y\rightarrow 0$, which is 
truly a throat, and $r=0$ is very similar to the horizon of the 
extreme Reissner-Nordstrom black hole. 
However, if $b <0$, then $r=0$ is at $y\rightarrow \infty$, and
we cannot speak of a horizon. Notice that this was the case where 
the potential
barriers were infinite. For $b =0$ the above change of
variables is not valid, but in this case the geometry is easily seen
to be 
${\bf R}^{p+2}\times S^{n+1}$, the size of the spheres being 
constant. The flat nature of the throat is another way of seeing that
the potential is zero for this case.

{}For these regular throats, the 
absorption cross section (when $b >0$) takes the form
\begin{equation}\label{regabs}
\sigma = \left({\omega\mu\over 2b}\right)^p {\pi 
\over \Gamma\left({p+1\over 2}\right)^2}
\Omega_{n+1} \mu^{n+1} \quad {\rm (regular\,\, throats)}.
\end{equation}
Notice that the dimensionful quantities that enter here are 
the area of the angular spheres, $\Omega_{n+1} \mu^{n+1}$, and the
`radius' of 
the $adS$ space, $\mu/b$. Thus we see that the low energy absorption 
cross section of {\it regular}
$p$-brane throats exhibits a sort of universality that generalizes
that of regular black holes.
Actually, as we will argue in subsec.\ \ref{pbr} below, 
the fact that $\sigma\propto\omega^p$ is suggestive of a 
simple worldvolume 
description of the absorption. For the particular cases of the 
D-threebrane $(D=10, p=3, b= 1)$, M-fivebrane
$(D=11, p=5, b=1/2)$, and M-membrane $(D=11, p=2, b=2)$, 
these absorption cross sections were
computed in \cite{threebr} \footnote{Eq.\ (\ref{regabs})
corrects the numerical factor in
\cite{threebr} for the case of the M-membrane.}.

When $\alpha <1$ the geometry is singular, but then it is conformal 
to one of the regular cases just described.  
The conformal factor, which 
must go like $r^{2\alpha-2}$ near $r=0$, can 
be physically meaningful
in many cases, as for example, when it has its origin in 
compactification from a higher dimensional metric. 
In such
cases the singularity at $r=0$ can be blown up and regularized
by revealing internal 
dimensions \cite{ght}. One can easily check that the wave equation 
for scattering in the reduced metric is 
the same as in the higher dimensional space. An example of this
will be presented in next section. The three different
classes of potential barriers $(b >,<,= 0)$ discussed above can be 
related to the conformal classes of the three 
regular throat geometries, i.e., $(adS)_{p+2}$ near $y=0$, 
$(adS)_{p+2}$ near $y=\infty$, 
and ${\bf R}^{p+2}$, all crossed with $S^{n+1}$.

One family of geometries of particular importance and which falls
into the
class of homogeneous branes described by (\ref{isotr}) 
is constituted by
the supergravity $p$-branes (see \cite{stelle}, and references 
therein). 
These arise as solutions of dilaton
gravity in $D=p+n+3$ spacetime dimensions, with a 
$(p+1)$-form gauge potential minimally 
coupled to gravity,
and dilaton coupling $a_d$. The metrics are
\begin{eqnarray}
ds^2 &=& H^{-{4n\over \Delta (p+n+1)}}\eta_{\mu\nu} dx^\mu dx^\nu
+ H^{{4(p+1)\over \Delta (p+n+1)}} (dr^2 + r^2 d\Omega_{n+1}),
\nonumber\\
H &=& 1 + \left({\mu\over r}\right)^n
\end{eqnarray}
where $\Delta = a_d^2+ 2(p+1)n/(p+n+1)$. Regular throats correspond
to non-dilatonic solutions, $a_d=0$. The D-$p$-branes of string theory
correspond to $D=10$ ($n=7-p$), $\Delta=4$, and among them, 
only the threebrane is nonsingular. In the notation of (\ref{isotr}),
all of them have $\alpha\leq 1$.

\section{Comparison with weakly coupled models of the 
absorption}\label{wkly}

In some cases, the absorption cross sections computed using the 
semiclassical methods of Sec.\ \ref{absor} can be compared with 
calculations at weak coupling (typically, at tree level) using a
microscopic model of the absorbing object. 

In the previous section we
have seen that absorption by regular surfaces presents particularly
simple features. As we will argue, these are as well the cases where
the correspondence with a microscopic description seems possible.
Since most absorption surfaces are singular,  
they should be somehow 
resolved if we want to compare the results of macroscopic 
and microscopic
calculations. In the following we discuss two possible resolutions
of the singularities. The first is, as mentioned above and discussed
in more detail in \cite{ght}, that the singularity is an effect
of longitudinal contraction along a compact direction, so that a
regular description is possible in higher dimensions. An example of
this is presented in \ref{effst}. In the other two subsections
the regularization will invoke, in the spirit of \cite{sen2}, 
short distance stringy effects near the singularity.
Two sorts of regular effective geometries are envisaged: the
stretched horizon, for degenerate states, and the stretched throat
for non degenerate ones.

\subsection{Absorption by the ground state of effective strings}
\label{effst}

The low energy dynamics of $D=5$ black holes with three charges,
and of $D=4$
black holes with four charges have been shown to be remarkably well
reproduced by microscopic models of ``effective strings''. 
These arise as 
$(1+1)$-dimensional conformal field theories at the string-like
intersection of some bound states of
branes: D-strings and D-fivebranes for the $D=5$ black hole 
\cite{ello}, and a (less well understood) triple intersection 
of M-fivebranes for the $D=4$ black hole \cite{kletse}. In addition,
one of the charges
is provided by momentum running along the intersection string. The
low energy absorption cross sections \cite{dmtwo,gubkleb1} 
and greybody factors \cite{ms,gubkleb2} for minimally coupled scalars
have been shown to be correctly reproduced in certain regions of
parameter space. 

The $D=5$ black hole with three charges and the $D=4$
black hole with four charges are particularly selected because they
possess a regular horizon of finite size in the extreme limit. 
The area of this horizon correctly reproduces the degeneracy of 
BPS states of the effective string for large charges. 
Essentially, this degeneracy
is due to the large number of states that can carry any
specified amount of momentum along the string. Here, instead,
we would like to
consider {\it non-degenerate} states, i.e., the ground states 
in which the
effective string is extremal (BPS) {\it and} 
has zero momentum number. 
The entropy of these ground 
states is consistently zero:
in $D=5,4$ these are black
holes with two and three charges, respectively, and have singular 
horizons of zero area.

The singularities, however, disappear when the solutions are
lifted to strings in $D=6,5$. 
Consider, for definiteness, the ground state of the $D=6$ effective 
string (things work very similarly for the $D=5$ effective 
string). The metric is
\begin{eqnarray}
ds^2 &=& H^{-1/2}(-dt^2 +dz^2) + H^{1/2} 
(dr^2 +r^2 d\Omega_3),
\nonumber\\
H&=& \left(1 + {r_{1}^2\over r^2}\right)
\left(1 + {r_{5}^2\over r^2}\right).
\end{eqnarray}
This geometry has a regular throat at $r=0$.
The absorption of low energy scalars follows easily from 
(\ref{theeq}) or (\ref{regabs}), 
\begin{equation}\label{sigeff1}
\sigma = \pi^3 \omega (r_1 r_5)^2.
\end{equation}

This is to be compared with the calculation for the same process 
in the weakly coupled description. Here, the absorption of a 
minimally coupled scalar by the ground
state of the effective string is described in terms of
interactions in the worldsheet, and can be readily obtained 
from the results in \cite{dmtwo,dmthree},
\begin{equation}\label{sigeff2}
\sigma = {\pi^3 g^2 \over V_4} \omega Q_1 Q_5,
\end{equation}
where $g$ is the string coupling, $(2\pi)^4 V_4$ 
is the volume of four compact (internal) dimensions, and 
$Q_1$, $Q_5$ are integer normalized D-string and D-fivebrane 
charges. Using 
the relation between these quantities and the black hole parameters 
$r_1$, $r_5$, \cite{ms}, the two results (\ref{sigeff1}) and 
(\ref{sigeff2}) agree precisely.

This problem can be regarded as a particular case of
\cite{ms,gubkleb2}, although a rather extreme one.
The fact is that the classical result for $\sigma$, 
Eq.\ (\ref{sigeff1}), 
can be obtained as a limit of the calculation in \cite{ms}, 
even if in that paper one assumed, in principle, a regular
black hole horizon. But it turns out that the validity of the 
solutions in \cite{ms,gubkleb2}
extends to the region of parameters of non-degenerate
states where the black hole 
geometry is singular. The reason we have studied it 
here is to stress the point that,
even if a geometry may be singular (like the $D=5$ extreme 
black hole with two charges), the low energy absorption rates 
computed from it can still be meaningful, particularly
if the solution can be lifted to a higher dimensional regular
geometry. The same conclusion could be drawn from \cite{threebr}.

\subsection{BPS fundamental strings: the stretched horizon and
stretched throat}\label{bpsfs}

There are essentially two kinds of (perturbative) BPS states of a 
fundamental string on a circle: 

\begin{itemize}

\item Degenerate states,
where oscillators of only one chirality are excited, say, $N_R=0$, 
$N_L\neq 0$. For large $N_L$ the logarithm of the 
degeneracy of such states
is $\sim \sqrt{N_L}$, or, using the level matching condition,
$\sim \sqrt{n_p n_w}$. 

\item Non-degenerate, or ground, states, with no
massive oscillations, $N_L=N_R=0$. 
In this case, the string winds around the circle, but does not carry 
momentum ($n_p=0$) \footnote{The case $n_w=0$, $n_p\neq 0$ 
is $T$-dual to this one.}.

\end{itemize}

These perturbative BPS states have been argued
\cite{duff,sen2} to correspond to extremal black holes
with electric NS-NS charges.
Such black holes can be viewed as the result of reducing along 
the string 
the solution describing
a string with winding number $n_w$ and momentum number $n_p$. 
In $D=n+4$ spacetime dimensions, the Einstein frame metric 
for this is
\begin{eqnarray}\label{string}
ds^2 &=& f_w^{-{n\over n+2}}\left[ -dt^2 + dz^2 +{r_p^n\over r^n}
(dt-dz)^2\right] + f_w^{2\over n+2}(dr^2 +r^2d\Omega_{n+1}),
\nonumber\\
f_{w} &=& 1+{r_{w}^n\over r^n}.
\end{eqnarray}
The coordinate $z$ will be thought of as being
periodically identified 
on a circle with 
radius $R$. Then the integer normalized charges corresponding 
to winding and 
momentum numbers are
\begin{equation}\label{intchrg}
n_w = {n\Omega_{n+1}\over (2\pi)^n}
{V_{(10-D)}\over g^2} r_w^n,\qquad
n_p = {n\Omega_{n+1}\over (2\pi)^n}
{V_{(10-D)}\over R^2 g^2} r_p^n.
\end{equation}
We assume that there are $10-D$ dimensions (not explicit 
in (\ref{string})) compactified in a volume 
$(2\pi)^{10-D}V_{(10-D)}$, and we have set 
$\alpha'=1$.

In this geometry $r=0$ is a singular point. Eq.\ (\ref{theeq}) 
yields an absorption cross section for low energy scalars,
\begin{equation}\label{wrong}
\sigma \sim (r_p^n r_w^n)^{n\over 2(n-1)} \omega^{1\over n-1}, 
\end{equation}
when $r_p\neq 0$, and 
\begin{equation}\label{wrong2}
\sigma \sim (r_w^n)^{n\over n-2} \omega^{n+2\over n-2}, 
\end{equation}
when $r_p= 0$. 
These show a marked dependence on the number of non-compact dimensions
in which the string lives. This is in contrast with the absorption of 
scalars, such as dilatons, as calculated in 
perturbative string theory. 
To this effect, we can use the amplitudes for dilaton scattering 
computed in
\cite{dmthree}. For the absorption by
degenerate BPS states, i.e., those 
having non-zero momentum number $n_p$, one finds
\begin{equation}\label{degabs}
\sigma \sim {g^2\sqrt{n_p n_w}\over R V_{(10-D)}},
\end{equation}
whereas if the string is in its ground state ($n_p=0$),
\begin{equation}\label{nondegabs}
\sigma \sim  {g^2\omega n_w\over V_{(10-D)}}.
\end{equation}
As could be expected, the perturbative absorption of dilatons is  
(apart from the trivial factor of the compact volume, and
numerical factors of order one that we omit) essentially independent
of the number of non-compact spatial directions. The scalings
of charges and frequencies {\it never} agree 
with the classical results
(\ref{wrong}), (\ref{wrong2}).

Indeed, we would not have expected a calculation in the geometry 
(\ref{string})
to yield the correct result, since it does not even lead to the 
correct
entropy for degenerate states with $n_p,\,n_w\neq 0$. Sen \cite{sen2} 
has argued that, 
since the black hole with these charges has a singularity at $r=0$, 
stringy effects should come into play at a string length distance. 
These would regularize the geometry so that a stringy 
{\it stretched horizon} appears, 
whose area reproduces the degeneracy of string states. 

In the present problem, it is not only the geometry
what might be altered by $O(\alpha')$ effects, but it can also
happen that non-minimal couplings of the scalar to the metric become 
important near the absorption surface. Thus it is better to think 
of the geometry with the stretched
horizon as an {\it effective} geometry for the propagation of the
scalar.

{}For the absorption by a degenerate BPS state it is easy to see that 
by simply placing the absorption surface at a distance $r\sim 1$,
then the absorption cross section gets corrected to the right
value (\ref{degabs}), up to factors of order one. Indeed, such an
absorption surface has a finite area, and according to 
(\ref{reghor}), this is equal to the absorption cross
section,
\begin{equation}
\sigma_{stretch}= A_{stretch} \sim \Omega_{n+1} (r_w r_p)^{n/2}.
\end{equation}
This reproduces the precise scaling of charges and frequency (as well
as couplings and moduli)
in (\ref{degabs}). This agreement can be seen as a consequence
of combining two facts: first, the absorption cross 
section by strings at low energy is, quite generally,
proportional to the degeneracy of the string state \cite{stanford},
and, second, the area of the stretched horizon yields the correct 
degeneracy of BPS strings \cite{sen2}.

So much for degenerate states, with non zero entropy. Things are not
so straightforward for ground states, since, even if their
geometries are singular as well at $r=0$,  they are not supposed to
generate a regular stretched horizon of finite area. Doing this would 
incorrectly assign them a non zero entropy. On the other hand, 
if the effective geometry near the string
is going to be a regular one, then our previous considerations
suggest that, instead of a regular horizon, we should
expect the string to develop a regular {\it stretched throat}.

We will implement this proposal with a minimal number of assumptions. 
Specifically, we will consider that the absorption surface is at 
$\bar r\equiv r-r_0=0$, with $r_0\sim 1$, and that the geometry 
still takes a homogeneous form. But now, instead of $f(r) =h(r)= 
f_w^{2\over n+2}(r)$, 
we take the metric near $r\approx 0$ to be a regular surface, 
\begin{eqnarray}
ds^2 &\approx& \gamma_{tt}(dt^2-dz^2) + 
{f_w^{2\over n+2}(r_0)\over \bar r^2}(d\bar r^2 +\bar r^2 
d\Omega_{n+1})
\nonumber\\
&\approx& 
\gamma_{tt}(dt^2-dz^2) + {(r_w^n)^{2\over n+2}\over \bar r^2}
(d\bar r^2 +\bar r^2 d\Omega_{n+1}).
\end{eqnarray}
Notice that, when fixing the form for $f(\bar r) = h(\bar r)$, 
we have not introduced
any new arbitrary factor of $r_w$. This is required by the fact 
that the throat should appear at distance $\sim 1$.
Nor have we assumed any particular form for
the metric $\gamma_{\mu\nu}$ in the string directions, 
apart from keeping it homogeneous. Now we find
\begin{equation}
U\; \sqrt{-\gamma^{tt} f} \approx {r_w^n\over \bar r},
\end{equation}
so that comparison with (\ref{parexp}) 
yields\footnote{The possibility that $a=b=0$
is discarded if we require the area of the absorption
surface to be zero.} $a=2b$ and 
$ST\approx r_w^n$. From 
(\ref{theeq}), the resulting absorption cross section is
\begin{equation}
\sigma \approx \omega r_w^n,
\end{equation}
which, using (\ref{intchrg}), agrees with the worldsheet calculation 
(\ref{nondegabs}) up to the numerical factor.

\subsection{Absorption by homogeneous $p$-branes}\label{pbr}

We saw in Sec.\ \ref{anat} that the absorption of the scalar
by a homogeneous $p$-brane with regular 
throat goes like $\sigma\propto\omega^p$. We now
follow \cite{threebr} to argue that
a simple microscopic coupling in the worldvolume of the $p$-brane 
reproduces this same behavior. Specifically, assume that the
$p$-brane worldvolume action contains a cubic coupling between the 
scalar $\Phi$ and a set of fields in the worldvolume, $X^i$,
\begin{equation}\label{wvcoup}
I_{int}
\sim \int d^{p+1} y\; \Phi(y)\; \partial_\alpha X^i\partial^\alpha X^i,
\end{equation}
where $y^\alpha$ ($\alpha =0,\dots,p$) are worldvolume coordinates.
The fields $X^i$ will in many cases be those that 
describe oscillations in some
directions transverse to the $p$-brane, but they could also be 
associated to longitudinal dynamics. In any case, the above coupling
is among the simplest and most widely
conceivable to appear in any worldvolume description.

The amplitude at tree level for the absorption of a quantum of $\Phi$
with energy $q^0=\omega$, and creation of a pair of modes of $X^i$ with
energies $k_1^0=k_2^0=\omega/2$ each, and opposite momenta $\vec{k}$, 
$-\vec{k}$, is
\begin{equation}
A\sim k_1\cdot k_2 = {\omega^2\over 4} - \vec{k}\cdot(-\vec{k}) = 
{\omega^2\over 2},
\end{equation}
and the absorption cross section,
\begin{eqnarray}
\sigma &\sim& \int d^p k_1 \int d^p k_2\; \delta^{p+1}(k_1+k_2 - q)
{|A|^2\over q^0 k_1^0 k_2^0}\nonumber\\
&\sim& \omega^p.
\end{eqnarray}
The last line follows in fact from simple scaling. 
Thus, it seems possible that
regular $p$-branes, regardless of whether or not
they are related to string
or M-theory, may admit a simple microscopic description
at weak coupling.

This sort of analysis, which is essentially based on power counting
of frequency and 
momenta, can be generalized, as has been recently 
done in \cite{das}, to include generic couplings, incorporating
fermions and higher dimension operators. This can be used
to identify further couplings that reproduce aspects
of the absorption rates. Let us just point out that, if the couplings 
are local, then it follows that the exponent of $\omega$ in the 
absorption cross section is always an integer. Since, in general,
the power of $\omega$ in (\ref{theeq}) is a fractional number, 
this would rule 
out the simplest correspondence between a large number of gravity 
$p$-branes and weakly coupled worldvolume theories.

In fact, for the $D$-branes of string theory we would expect to find
$O(\alpha')$ corrections similar to those 
that led us to introduce the stretched throat as an effective geometry
for the scattering. Such modification could
be thought of as arising from fluctuations of the open strings ending
on the D-brane. Again, the simplest alteration
assumes that
the metric along the worldvolume continues to be homogeneous, and
that the geometry near the the absorption surface at $\bar r=r-r_0 = 0$
is a non singular
stretched throat, $f(\bar r)=h(\bar r) \propto \bar r^{-2}$. 
Then we recover, 
as in (\ref{regabs}), $\sigma\propto \omega^p$. Just as with the 
BPS string, $\sigma$ is largely
independent of the geometry 
$\gamma_{\mu\nu}$ parallel to the $p$-brane. This is most
easily seen from the fact that (\ref{regabs}) does not 
depend on $\nu$, while $\beta=b$ affects only the numerical factor.

Things would be more complicated if we wanted to include 
higher partial
waves. For the class of geometries considered in the appendix, i.e.,
those with $a=n$, non singular throats correspond to having 
$b={n\over p+1}$. Then, using (\ref{theeql}),
\begin{equation}\label{lsiggr}
\sigma^l \propto \omega^{p + 2l {p+n+1\over n}}.
\end{equation}
On the other hand, in \cite{threebr} it is proposed that the 
relevant worldvolume couplings for the absorption of angular momentum
$l$ come from 
\begin{equation}
\int d^{p+1} y\; \Phi(X,y)\; \partial_\alpha X^i\partial^\alpha X^i
= \sum_l {1\over l!}
\int d^{p+1} y\; (\partial_{i_1}\dots\partial_{i_l}\Phi)
X^{i_1}\dots X^{i_l}\; \partial_\alpha X^i\partial^\alpha X^i.
\end{equation}
Such couplings would yield
\begin{equation}\label{lsigwc}
\sigma^l\propto \omega^{p+ l(p+1)},
\end{equation}
and we see that (\ref{lsiggr}) and (\ref{lsigwc}) agree only if
$2(p+1) = n(p-1)$. The branes considered in \cite{threebr} fall into 
this class. The situation for the effective strings has been analyzed
in \cite{ms2,matl,gubl}. In other situations it appears that
one may need further information about the correspondence with the
worldvolume theory to reach an agreement.

We have not specified any dependence on coupling constants or charges 
in this discussion. To analyze this point we should be more 
precise about, on the one hand, the normalization of states
and worldvolume interactions 
that may account for the absorption, and
on the other hand, the location of the stretched throat. For 
fundamental strings we took the latter to be at the radius where the
curvature (in string frame) and other fields take on values of string 
scale order, which is at $r\sim \sqrt{\alpha'}= 1$. For D-branes 
it need not be 
there, and it is probably at a much shorter distance. It remains to be
seen if there is any plausible definition of the radius of the
stretched throat for D-branes that leads to the correct scaling
with the charges.

\section{Conclusions}\label{concl}

We have managed to derive an expression, Eq.\ (\ref{theeq}), 
for the low energy absorption cross section of minimal scalars that
covers a very broad class of geometries of interest. For the 
particular
cases where the absorption surface is regular, either a
finite area surface, or a regular throat, 
the absorption depends 
only on geometrical parameters of the absorption surface: 
the total area in (\ref{reghor}), and the limiting 
sizes of the $(adS)_{p+2}$ 
and $S^{n+1}$ down the throat in (\ref{regabs}). 
This is the way in which the universal result
for black holes in \cite{dgm}
is generalized to regular extended objects.

The simple features of absorption by regular surfaces call for
a more detailed microscopic picture. However, by no means all the
geometries of branes of interest are non singular. This is in fact
an indication that
their singularities should be resolved in some way. We have discussed
two ways in which this can happen: one involves lifting the
geometry to higher
dimensions; the other calls upon short distance stringy effects. 
The latter only
suggest that some modification should happen near the singularity.
We have assumed, following \cite{sen2}, that the net result 
is to yield an effective regular geometry, and shown that this 
leads to non-trivial correct results.

A stretched horizon had
already been invoked to account for the entropy of 
degenerate string states. However, the entropy counting 
hardly tells anything
about what the geometry near the location of the string should be
for non-degenerate states. One only knows that
the area of the horizon should be kept as zero. 
This is where the absorption
cross sections are a useful tool for obtaining further information,
which has led us to the concept of a stretched
throat. For the ground state of fundamental strings this 
is unambiguously fixed by requiring that

(a) The geometry keeps its homogeneous form, and the
absorption surface is regular with zero area.

(b) The throat appears at string size scale, and so it
does not introduce new factors of the charges in the 
metric coefficients.

The stretched throat is also useful when dealing with other
homogeneous $p$-branes, which typically will be extreme objects.
If a simple worldvolume coupling to the scalar is to be responsible
for the absorption, then a regularized throat yields results that 
are compatible
at least in their dependence with the frequency. 
We hope that further work will lead to refinement and extension
of the results presented here.

\section*{Acknowledgements}
This work has been partially 
supported by an FPI fellowship (MEC-Spain) 
and by grant UPV 063.310-EB225/95.

\appendix
\section{Absorption of higher partial waves}

In this appendix we compute the absorption cross section for the
higher partial waves of a minimallly coupled scalar. 

The wave function is of the form $\Phi = e^{-i\omega t}
P_l (\cos\theta)\phi^l_\omega (r)$, so the radial wave equation reads
\begin{equation}\label{waveql}
\left[ -\omega^2 \gamma^{tt}(r) f(r) 
-{l(l+n)\over r^2}{f(r)\over h(r)}
+{\sqrt{-\gamma^{tt} f(r)}\over U(r)}
{d\over dr} {U(r)\over \sqrt{-\gamma^{tt} f(r)}}
{d\over dr}\right]\phi^l_\omega(r) =0.
\end{equation}
In order to be able to solve the wave equation we will
restrict to metrics with $f(r) = h(r)$ \footnote{It would also be
possible to solve for $f(r)/h(r)= {\rm constant}$.}. Furthermore, the 
sort of matching between far and near solutions that 
we will perform will
be possible, for generic $l$, only if in (\ref{parexp}) we have $a=n$. 

The fluxes are
computed using the same formula (\ref{flux}), and the absorption
cross section simply gets multiplied by a factor of the degeneracy
of the representation of the rotation group \cite{matl,gubl},
\begin{equation}
\sigma^l = {(2\pi)^{n+1}\over \omega^{n+1}\Omega_{n+1}} 
{{\cal F}_0\over {\cal F}^{in}_\infty}{2l+n\over n}{l+n-1\choose l},
\end{equation}

Introducing $r^*$ and $\psi$ as before, the potential in the 
Schr\"odinger equation (\ref{scheq}) is now
\begin{equation}
V_l = {1\over \sqrt{U}} {d^2 \sqrt{U}\over (dr^*)^2}
+{l(l+n)\over (-\gamma^{tt} f) r^2}.
\end{equation}
At large distances this is $V_l\approx {(n+2l)^2-1\over 4 r^2}$,
and we solve as 
\begin{equation}\label{farsoll}
\phi^{l,\infty}_\omega = {1\over (\omega r)^{n/2}} 
\left( A J_{n/2+l}(\omega r) +B J_{-(n/2+l)}(\omega r)\right).
\end{equation}
Close enough to $r=0$ the wave equation is again of Bessel type, and
is solved by
\begin{equation}\label{nearsoll}
\phi^{l,\rm near}_\omega  ={1\over (\omega r)^{n/2}}
H^{(1)}_{n+2l\over 2b} \left({\omega T\over b r^b}\right),
\end{equation}
where we have already set $a=n$.
Again, if there exists an overlap region,
the matching of far and near solutions can be performed 
with $B=0$. The fluxes take the same form as in (\ref{fluxinf}) and 
(\ref{fluxo}), and the final result can be written as
\begin{equation}\label{theeql}
\sigma^l = \pi \Omega_{n+1} 
\left[{\Gamma({n\over 2}+1)\over \Gamma({n\over 2}+l+1)
\Gamma({n+2l\over 2b})}\right]^2 {2l+n\over n}{l+n-1\choose l} S
\left({T\over b}\right)^{{n+2l\over 
b}-1}\left({\omega\over 2}\right)^{{n+2l\over 
b}+2l-1}.
\end{equation}

\end{document}